\documentclass[a4paper,11pt]{article}
\usepackage{jcappub} 
\usepackage{lineno}

\usepackage{amsmath} 
\usepackage{algorithmic}
\usepackage{bm}

\newenvironment{extdatafigure}{
  \renewcommand{\figurename}{Fig.}
  \begin{figure*}[!htbp]
}{
  \end{figure*}
  \renewcommand{\figurename}{Fig.}
}
\newcounter{algorithm}

\title{\boldmath Closing the Observational Gap in Cosmic Dynamics: AI-Enabled Reconstruction of the Universe’s Vorticity and Rotational Flow Morphology}

\author[a,b,c]{Ziyong Wu}
\author[a]{Xu Xiao}
\author[d]{Fuyu Dong}
\author[e]{Juhan Kim}
\author[f]{Yan-Chuan Cai}
\author[g]{Yang Wang}
\author[h,i,b]{Xi Kang}
\author[a,j]{Le Zhang}
\author[a,j]{Xin Wang}
\author[a,j]{Xiao-Dong Li}

\affiliation[a]{School of Physics and Astronomy, Sun Yat-Sen University, Guangzhou, China}
\affiliation[b]{Purple Mountain Observatory, Chinese Academy of Sciences, Nanjing, China}
\affiliation[c]{School of Astronomy and Space Sciences, University of Science and Technology of China, Hefei, China}
\affiliation[d]{South-Western Institute for Astronomy Research, Yunnan University, Kunming, China}
\affiliation[e]{Korea Institute of Advanced Studies (KIAS), Seoul, Republic of Korea}
\affiliation[f]{Institute for Astronomy, University of Edinburgh, Edinburgh, UK}
\affiliation[g]{Peng Cheng Laboratory, Shenzhen, China}
\affiliation[h]{Institute for Astronomy, The school of Physics, Zhejiang University, Hangzhou 310037, China}
\affiliation[i]{Center for Cosmology and Computational Astrophysics, Zhejiang University, Hangzhou 310027, China}
\affiliation[j]{CSST Science Center for the Guangdong–Hong Kong–Macau Greater Bay Area, SYSU, Zhuhai, China}

\abstract{The cosmic vorticity field—an essential tracer of nonlinear structure formation—has remained observationally inaccessible because transverse galaxy motions are difficult to measure and analytic models struggle to capture shell-crossing. Here we report an empirical reconstruction of this field by applying an artificial intelligence framework, trained on simulations of the concordance $\Lambda$CDM model, to Sloan Digital Sky Survey galaxies. The recovered three-dimensional velocity and vorticity fields reveal coherent vortical structures, including spiral flows in clusters, filaments, and voids, and the cosmic web inferred from vorticity closely matches that derived from density segmentation. The power spectra of the reconstructed velocity and vorticity fields agree statistically with $\Lambda$CDM predictions, and the inferred velocity field effectively removes redshift-space distortions, yielding an almost isotropic clustering signal. These converging lines of evidence, obtained from an independent perspective, reinforce the concordance cosmological model. By closing a long-standing observational gap, our results highlight the potential of AI-driven reconstruction to access otherwise unobservable quantities and to address fundamental questions in cosmology and galaxy formation.}

\begin{document}
\maketitle
\flushbottom

\section{Introduction} \label{sec:intro}

The large-scale structure (LSS) of the Universe provides a sensitive probe of cosmic evolution, illuminating fundamental physics such as the nature of dark matter and dark energy \cite{2009GReGr..41.1455M,1993ppc..book.....P} and the processes that govern galaxy formation \cite{2005Natur.435..629S,1978MNRAS.183..341W}. Peculiar velocities, generated by local gravitational potentials, deliver complementary dynamical information to galaxy number-density statistics. Measuring them, however, remains challenging. Radial components can be inferred from type Ia supernovae, Tully–Fisher, or fundamental-plane relations \cite{1993ApJ...413L.105P,SNIaflow...1997ApJ...488L...1R,SNIaflow...2004MNRAS.355.1378R,SNIaflow...2012MNRAS.420..447T,SNIaflow...2016ApJ...827...60M,TullyFisher...1977A&A....54..661T,TullyFisher...2006ApJ...653..861M,TullyFisher...2008AJ....135.1738M,FundPlan...1987ApJ...313...42D,FundPlan...1987ApJ...313...59D,FundPlan...2007ApJS..172..599S}, enabling tests of redshift-space distortions, the Alcock–Paczynski effect, the kSZ signal, and the ISW effect \cite{jackson1972critique,kaiser1987clustering,ap,Li2014,Li2015,Li2016,KR2018,1972CoASP...4..173S,1980MNRAS.190..413S,1967ApJ...147...73S,1968Natur.217..511R,1996PhRvL..76..575C}. The transverse components, by contrast, remain effectively unobservable.

Velocity-field reconstructions generally rely on the continuity equation relating density and velocity, achieving success on large, quasi-linear scales \citep{1995ApJ...441..449G,1995PhR...261..271S}. Such approaches, however, do not capture vorticity, which emerges through shell-crossing in the nonlinear regime \citep{2015MNRAS.454.3920H,2013ApJ...766L..15L}. Perturbative treatments break down precisely where vorticity becomes significant, and statistical estimators retrieve at most partial information, falling short of reconstructing the full field \citep{2000MNRAS.316..464K,2008MNRAS.389..497K,VelocityRecon...1991ApJ...379....6N,VelocityRecon...1992ApJ...390L..61B,1994ApJ...421L...1N,VelocityRecon...1995ApJ...449..446Z,VelocityRecon...1997MNRAS.285..793C,VelocityRecon...1999MNRAS.309..543B,VelocityRecon...2000MNRAS.316..464K,VelocityRecon...2002MNRAS.335...53B,VelocityRecon...2005ApJ...635L.113M,VelocityRecon...2008MNRAS.383.1292L,VelocityRecon...2008MNRAS.391.1796B,VelocityRecon...2012MNRAS.425.2422K,VelocityRecon...2012MNRAS.420.1809W,VelocityRecon...2015MNRAS.449.3407J,2017MNRAS.467.3993A,2019A&A...625A..64J}. Constrained simulations offer an alternative but remain computationally prohibitive \citep{2013MNRAS.432..894J,2015MNRAS.454.3920H,2009ApJ...705..156H,2016ApJ...831..164W}. Consequently, despite decades of effort, a large-volume three-dimensional observational determination of the cosmic vorticity field has not been achieved.
These limitations motivate the reconstruction strategy introduced below, which delivers access to the missing transverse information using a data-driven approach.

In this work, we employ a deep neural network to reconstruct the vorticity field implied by the peculiar velocities of SDSS galaxies. Simulation experiments demonstrate that UNet-based architectures can recover nonlinear velocity fields, including vorticity signatures down to $k \sim 1~h^{-1}\mathrm{Mpc}$ \cite{2021ApJ...913....2W,2023MNRAS.522.4748W,2024MNRAS.532.3947C,2024ApJ...969...76W}, whereas standard perturbation theory fails in this regime \cite{1980lssu.book.....P,2002PhR...367....1B,Pueblas:2008uv,Umeh:2023lbc}.

\section{METHODOLOGY} \label{sec:method}

\subsection{Data}

The observational foundation of this study is based on the Korea Institute for Advanced Study Value-Added Galaxy Catalog (KIAS-VAGC; \cite{2010JKAS...43..191C}), an extension of the New York University Value-Added Galaxy Catalog (NYU-VAGC; \cite{2005AJ....129.2562B}) compiled from the SDSS-I/II DR7 main galaxy sample \cite{2009ApJS..182..543A}. By incorporating redshifts from multiple auxiliary surveys, the KIAS-VAGC improves completeness and covers roughly $4000\,\mathrm{deg}^2$ of sky.

Training of the deep-learning model relies on mock SDSS DR7 catalogues constructed by \cite{2023ApJ...953...98D} from the Horizon Run 4 (HR4) $N$-body simulation \cite{2015JKAS...48..213K}, which adopts WMAP5 cosmological parameters $(\Omega_{\Lambda}, \Omega_m, \Omega_b, h, \sigma_8, n_s) = (0.74, 0.26, 0.044, 0.72, 0.794, 0.96)$. In each mock, the most bound dark-matter particle of a halo is designated as a galaxy, and real-space positions are mapped into redshift space within a spherical shell that emulates the SDSS footprint.

The authors of \cite{2023ApJ...953...98D} reproduced the observed selection function using subhalo abundance matching (SHAM) \cite{2004MNRAS.353..189V}, calibrated to the SDSS mass-selection function. By varying the observer’s position, they generated 108 independent, volume-limited mocks that track the DR7 main sample between $0.025 \leq z \leq 0.163$. The adopted magnitude cut, $M_r < -21.07 + 5 \log h$ (roughly $M_\star \sim 10^{11}\,\mathrm{M}_{\odot}$), yields a number density $\bar{n}_{\mathrm{gal}} \approx 10^{-3}\,(\mathrm{Mpc}/h)^{-3}$. We use 98 mocks for training, 5 for validation, and 5 for testing.

\begin{extdatafigure}
\centering
\includegraphics[width=0.9\textwidth]{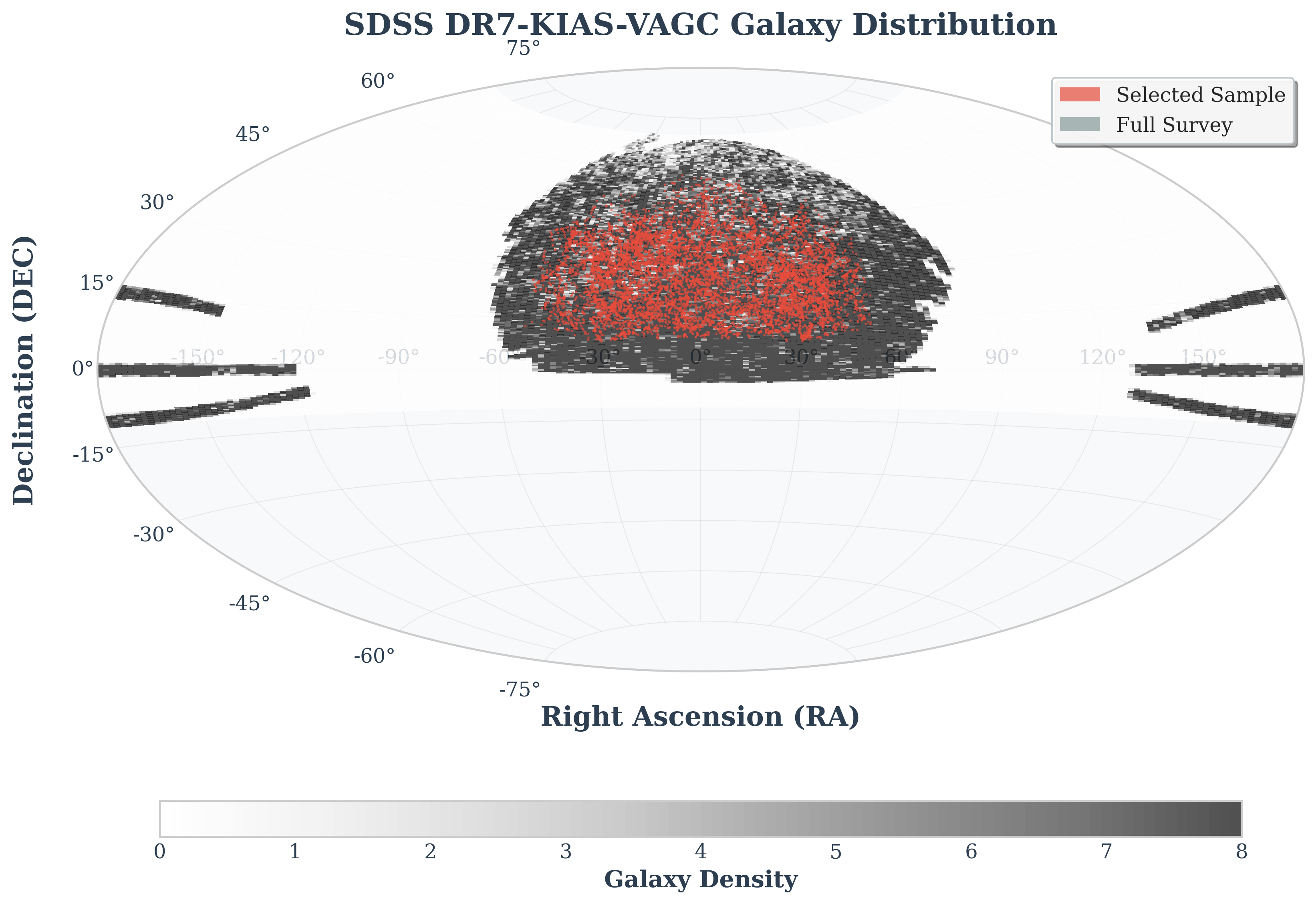}
\caption{SDSS DR7-KIAS-VAGC galaxy distribution in celestial coordinates. Aitoff projection showing spatial distribution of galaxies from SDSS DR7 survey. Background density map (grey scale) represents full survey coverage; red points indicate selected sample for cosmic web reconstruction. Sample filtered for redshift range $0.01 < z < 0.17$ and specific spatial coordinates, corresponding to two adjacent cubic volumes. Right ascension (RA) and declination (DEC) coordinates displayed on axes.}
\label{fig:app_fig1}
\end{extdatafigure}

For compatibility with convolutional neural networks, which operate on regular grids, galaxy positions are transformed from sky coordinates to comoving Cartesian coordinates. We select a high-density subregion within the survey volume defined by $x \in [-322, -118]$, $y \in [-204, 204]$, and $z \in [33, 237]$ in $h^{-1}\mathrm{Mpc}$. Fig.~\ref{fig:app_fig1} highlights this reconstructed domain in red against the full KIAS-VAGC catalogue in grey. The volume is sufficiently large to encompass coherent structures while limiting edge effects that could bias the inferred velocity and vorticity fields.

Owing to the approximate symmetry about the $y$-axis, the region can be partitioned into two mirrored sub-volumes. Each mock is therefore divided into two cubes of side length $204\,h^{-1}\mathrm{Mpc}$, corresponding to the $y>0$ and $y<0$ halves (denoted $y+$ and $y-$). These cubes have an average galaxy density of $6 \times 10^{-4}\,(h^{-1}\mathrm{Mpc})^{-3}$, which is adequate for training while preserving spatial uniformity. This step effectively doubles the number of training examples—yielding 196 cubes for training, 10 for validation, and 10 for testing—and enforces a consistent line-of-sight configuration, thereby minimising orientation-dependent biases.

\subsection{Data Pre-processing}
The pre-processing pipeline comprises three principal steps:

\begin{enumerate}
\item[1)]
For each mock we construct density and velocity fields on $(4h^{-1}\mathrm{Mpc})^3$ cells, yielding $51^3$ meshes when using the cloud-in-cell (CIC) interpolation. The density grids serve as network inputs and the velocity grids as training targets. Because CIC can underestimate clustering in low-density regions, potentially biasing two-point statistics, we also build fields with the Delaunay Tessellation Field Estimator (DTFE). To enforce a consistent line of sight, the fields are mirrored during pre-processing to align the $y+$ and $y-$ blocks. Gaussian smoothing with $(4\,h^{-1}\,\mathrm{Mpc})^3$ for density and $(8\,h^{-1}\,\mathrm{Mpc})^3$ for velocity is applied to the DTFE fields to suppress numerical noise.

\item[2)]
Following \cite{2023MNRAS.522.4748W}, we decompose each velocity vector into magnitude and direction rather than predicting the full vector directly. The network outputs these components and they are recombined to recover the velocity field. To mitigate boundary artefacts, the loss is evaluated only within the central $45^3$ cells.

\item[3)]
Given the simulation volume, no auxiliary linear velocity component is required. Prior to ingestion, density and velocity fields are normalised by constant factors $c$: for CIC fields we use $c=50$ (density) and $c=100$ (velocity), while DTFE fields adopt $c=1$ and $c=1000$, respectively.
\end{enumerate}

\begin{extdatafigure}
\centering
\includegraphics[width=\textwidth]{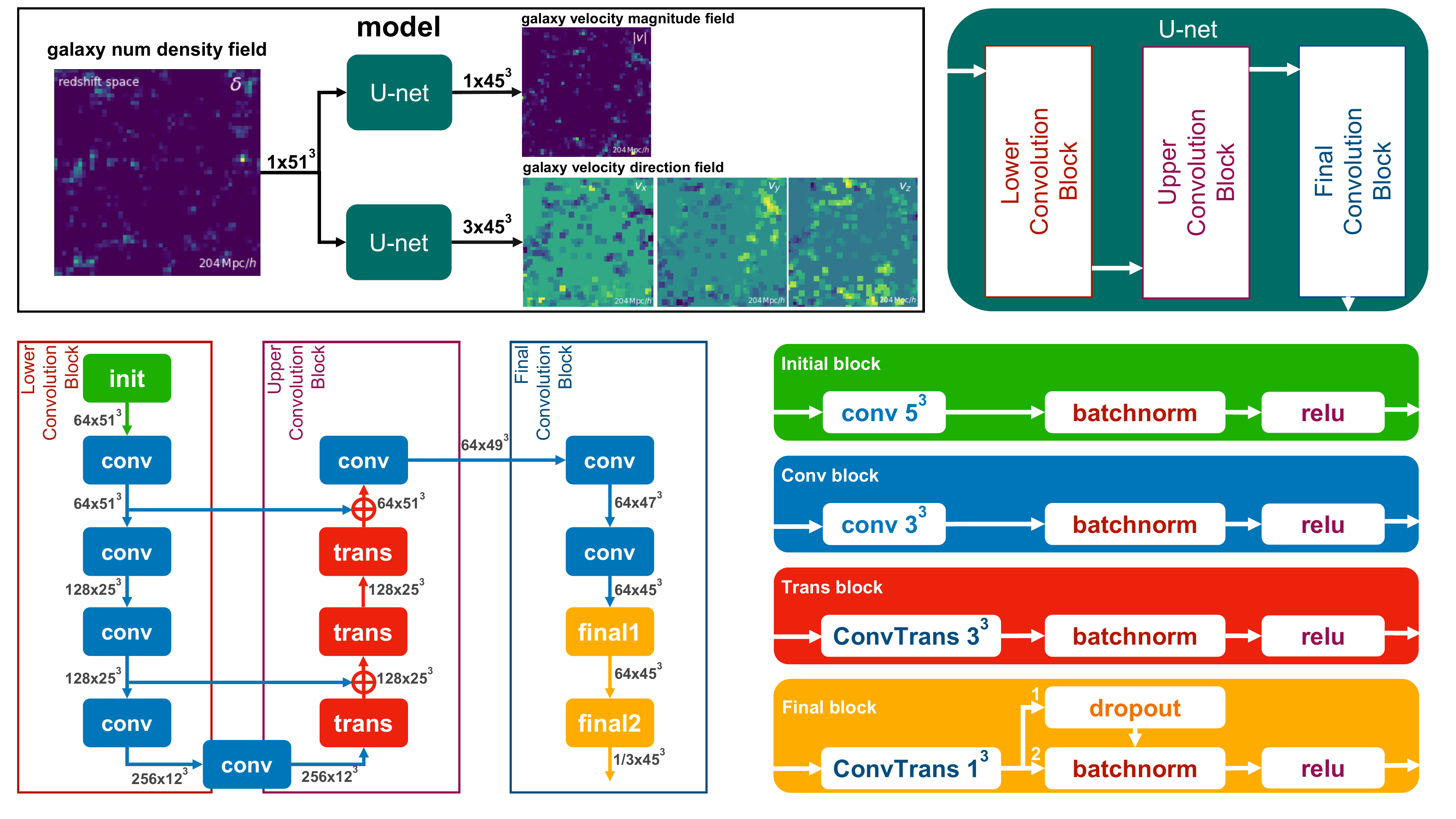}
\caption{The U-Net model used in this study builds upon the architecture from \cite{2023MNRAS.522.4748W}, with adaptations to enhance velocity reconstruction. The network takes a $51^3$-voxel input representing the redshift-space galaxy number density field (side length: $204\,\mathrm{Mpc}/h$) and employs a dual-path structure: one branch reconstructs the velocity magnitude ($|v|$), and the other the three velocity components ($v_x$, $v_y$, $v_z$), as illustrated in the upper-left panel. The model consists of upper, lower, and final convolution blocks, producing an output of dimension $45^3$ corresponding to a physical volume of $180^3\,(h^{-1}\mathrm{Mpc})^3$. The lower-right section details the three-block UNet structure, with layer specifications ('init', 'conv', 'trans', 'output') provided in the lower-left. A dropout rate of 0.3 is applied in the final block to improve performance and mitigate overfitting.}
\label{fig:app_fig2}
\end{extdatafigure}

\subsection{{Neural Network Architecture} }
The network architecture, illustrated in Fig.~\ref{fig:app_fig2}, extends the framework of \cite{2023MNRAS.522.4748W}. The input is the galaxy number-density field, and the output decomposes the velocity into magnitude and directional components. Two structurally similar sub-networks predict these quantities separately, yielding four output channels: the three directional components ($\hat{v}_x, \hat{v}_y, \hat{v}_z$) and the scalar magnitude. The complete velocity vector is recovered by combining these outputs. Relative to \cite{2023MNRAS.522.4748W}, we reduce the number of input channels from six to one because the galaxy number density in our sample does not support further subdivision, and we refrain from splitting the velocity into high- and low-speed components, as no performance gains were observed.

Fig.~\ref{fig:app_fig2} details the UNet topology. Coloured blocks denote operations linked by arrows, and annotations specify the channel counts, spatial dimensions, and convolution-kernel sizes. Key architectural elements include (i) stride-two convolutions to downsample or upsample the field, (ii) initial convolutional layers that enlarge the receptive field and accelerate the learning of large-scale features, (iii) output layers followed by dropout to control overfitting and adjust channel dimensionality, and (iv) batch normalisation coupled with ReLU activation to stabilise and non-linearise training.

Once trained, the UNet maps the redshift-space density field to the peculiar velocity field, enabling direct computation of derived statistics. The architecture comprises three main convolutional blocks—lower (red), upper (pink), and final (blue)—that progressively refine the representation. Because the simulation volume already captures the relevant large-scale dynamics, no auxiliary linear velocity component is included. To suppress boundary artefacts, the final block predicts velocities only within the central region of the density cube rather than across the full domain.

\subsection{Loss Function}

\begin{extdatafigure}
\centering
\includegraphics[width=0.8\textwidth]{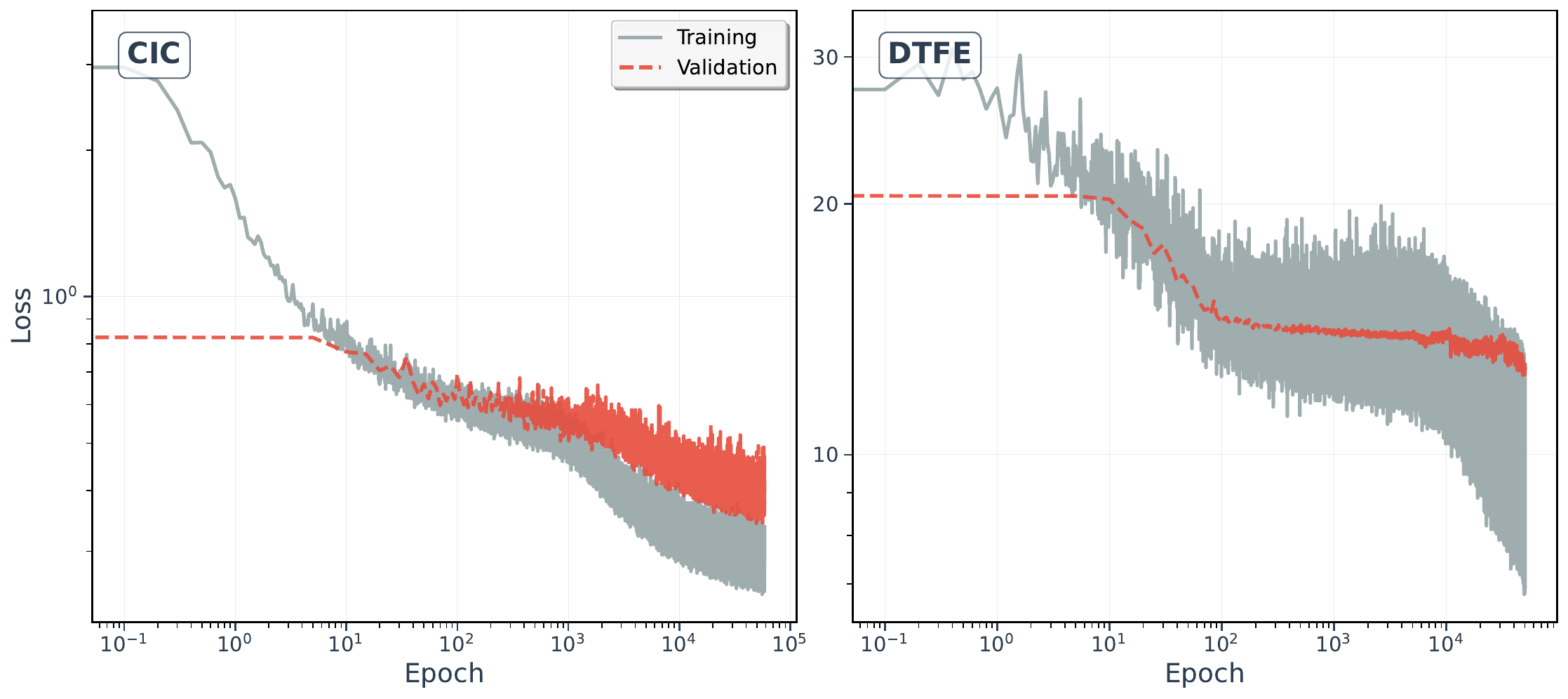}
\caption{Training and validation loss curves for UNet models. Two UNet models trained on different preprocessing methods: CIC (Cloud-in-Cell) and DTFE (Delaunay Tessellation Field Estimator). Both show training loss (solid lines) and validation loss (dashed lines) vs. epochs on linear (logarithmic) scales. }
\label{fig:app_fig3}
\end{extdatafigure}

Training seeks to minimise a loss that measures the discrepancy between the predicted velocity field $\bm{v}$ and the simulated truth $\bm{v}^{\rm true}$ at each voxel. To capture errors in both magnitude ($v \equiv |\bm{v}|$) and direction ($\hat{\bm{v}} \equiv \bm{v}/v$), we adopt a composite loss with three contributions:

\begin{equation}
\begin{aligned}
\mathcal{L} = \frac{1}{N} \sum_{n=1}^N \Bigg[
&\frac{1}{4} (v_n - v^{\rm true}_n)^2  \\
&+ \frac{1}{4}\sum_{i=1}^3 \left(1 - \cos\phi_i \right)  \\
&+ \frac{1}{4}\sum_{i=1}^3\sum_{j=1}^3 \frac{1}{9}
\left( \frac{d_i v_n}{dx_j} - \frac{d_i v^{\rm true}_n}{dx_j} \right)^2
\Bigg]
\end{aligned}
\label{eq:loss}
\end{equation}

Here $\cos\phi_i \equiv \hat{\bm{v}}_i \cdot \hat{\bm{v}}^{\rm true}_i$, $i$ indexes voxels, and $N$ is the total voxel count. The first term is a mean-squared error on the velocity magnitude, consistent with a Gaussian likelihood. The second term enforces directional agreement via cosine similarity, with weights reflecting the relative degrees of freedom (one for magnitude, three for direction). The third term penalises gradients to improve the reconstruction of divergence and vorticity. Coefficients are chosen empirically to ensure stable training. The loss is optimised with Adam \cite{2014arXiv1412.6980K}, and convergence is typically achieved after $\sim 10^4$ epochs (Fig.~\ref{fig:app_fig3}).

\begin{figure*}
\centering
\includegraphics[width=1\textwidth]{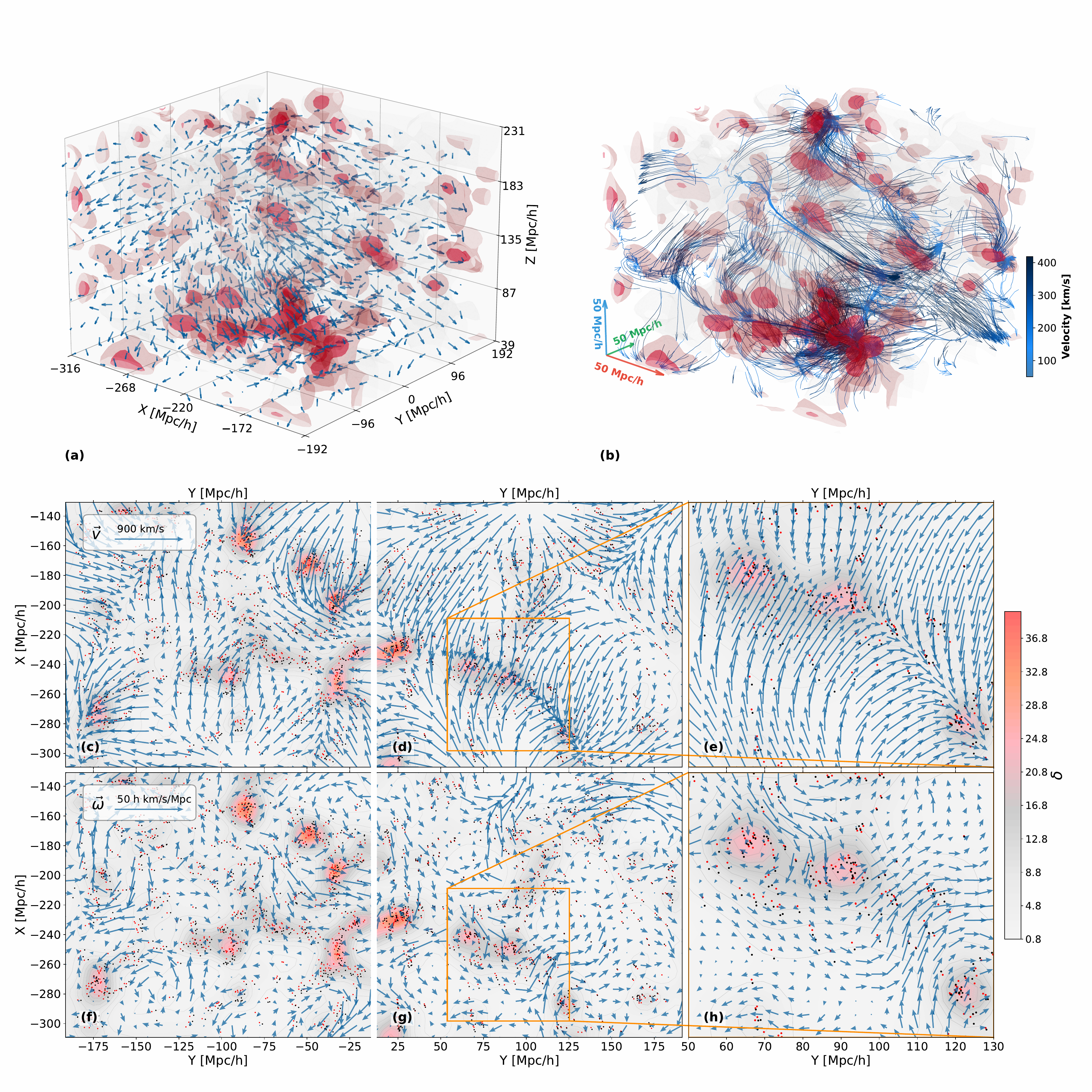}
\caption{Reconstructed velocity and vorticity fields from SDSS data. Top row shows 3D velocity field ($204\times408\times204 h^{-3}\rm{Mpc}^3$) with density isosurfaces and velocity vectors (a) and streamlines (b). Bottom row displays 2D slices ($204\times204\times24 h^{-3}\rm{Mpc}^3$) of velocity (c,d) and vorticity (f,h) fields with galaxy positions (red: real-space, black: redshift-space). Panels (e) and (h) show zoomed views of "SDSS Great Wall" region. Density fields use DTFE method; velocity fields reconstructed by U-Net AI model.}
\label{fig:fig1} 
\end{figure*}

\section{Result} \label{sec:result}

The reconstructed three-dimensional velocity field is shown in Fig.~\ref{fig:fig1}. Streamlines superposed on the galaxy density field reveal that the SDSS Great Wall—one of the survey’s most prominent structures—is accurately recovered. A zoomed $204\times408\times20\,(\mathrm{Mpc}/h)^{3}$ subvolume displays the velocity and vorticity fields derived with DTFE interpolation, while Fig.~\ref{fig:app_fig6} confirms that the CIC scheme yields analogous results. Both interpolation methods depict matter converging from underdense regions toward the Great Wall’s filaments and nodes, with the resulting flows generating pronounced vorticity near clusters and at pathway intersections. These vortical structures demonstrate how nonlinear gravitational evolution imprints rotation on initially irrotational flows, exposing the complex kinematics of the cosmic web.

\begin{extdatafigure}
\centering
\includegraphics[width=0.9\textwidth]{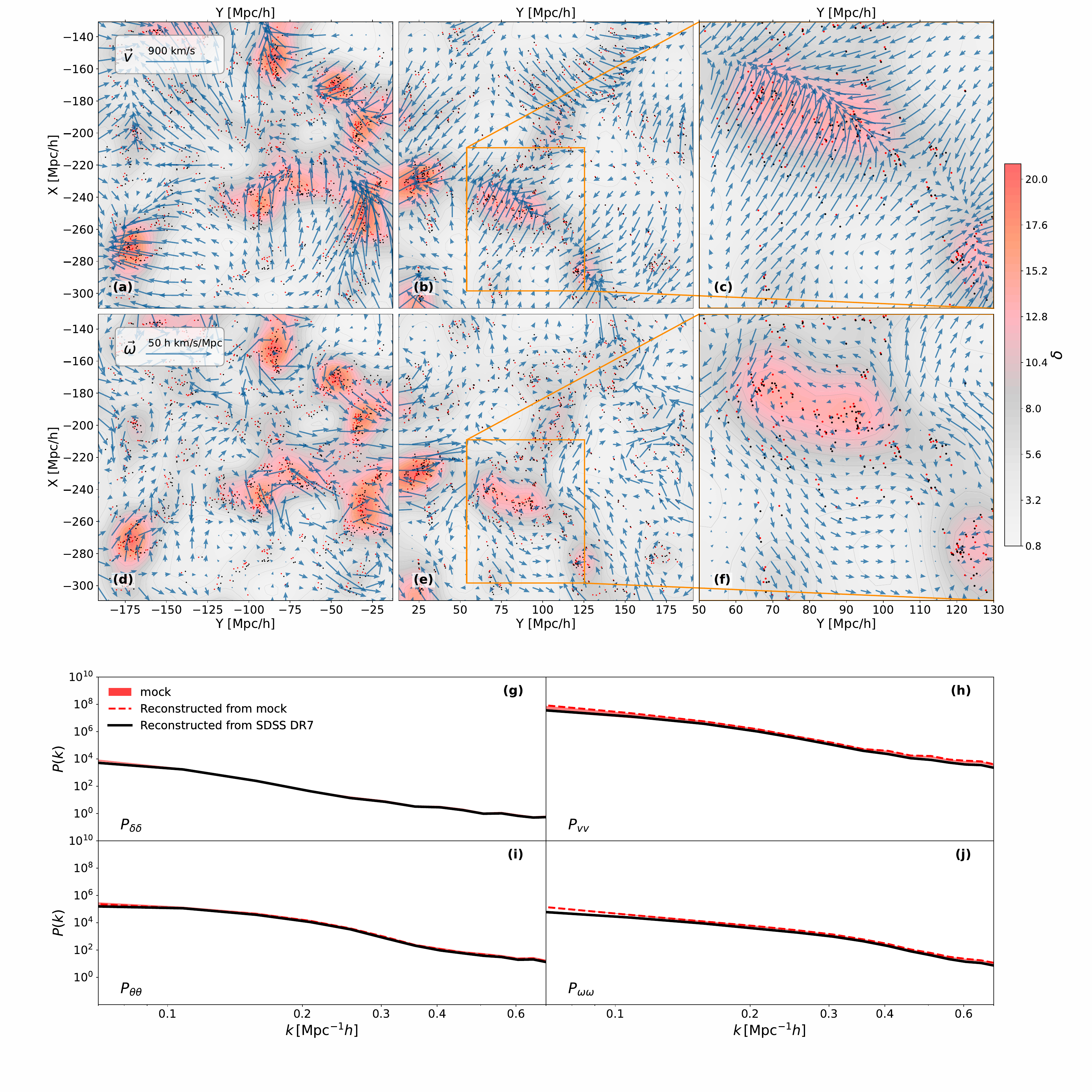}
\caption{Combined 2D slice visualization and power spectrum analysis for the results based on CIC preprocessing. Top row shows 2D slices of velocity (a,b) and vorticity (c,d) fields from SDSS DR7 data, with density contours and galaxy positions (red: real-space, black: redshift-space). Panels (e,f) provide zoomed views of highlighted regions. Bottom row displays power spectra: (g) $P_{\delta\delta}$, (h) $P_{vv}$, (i) $P_{\theta\theta}$, and (j) $P_{\omega\omega}$, comparing mock data (blue shaded), mock reconstructions (blue dashed), and SDSS DR7 reconstructions (black solid).}
\label{fig:app_fig6} 
\end{extdatafigure}

\subsection{Consistency with the $\Lambda$CDM framework}

\begin{figure*}
\centering
\includegraphics[width=0.8\textwidth]{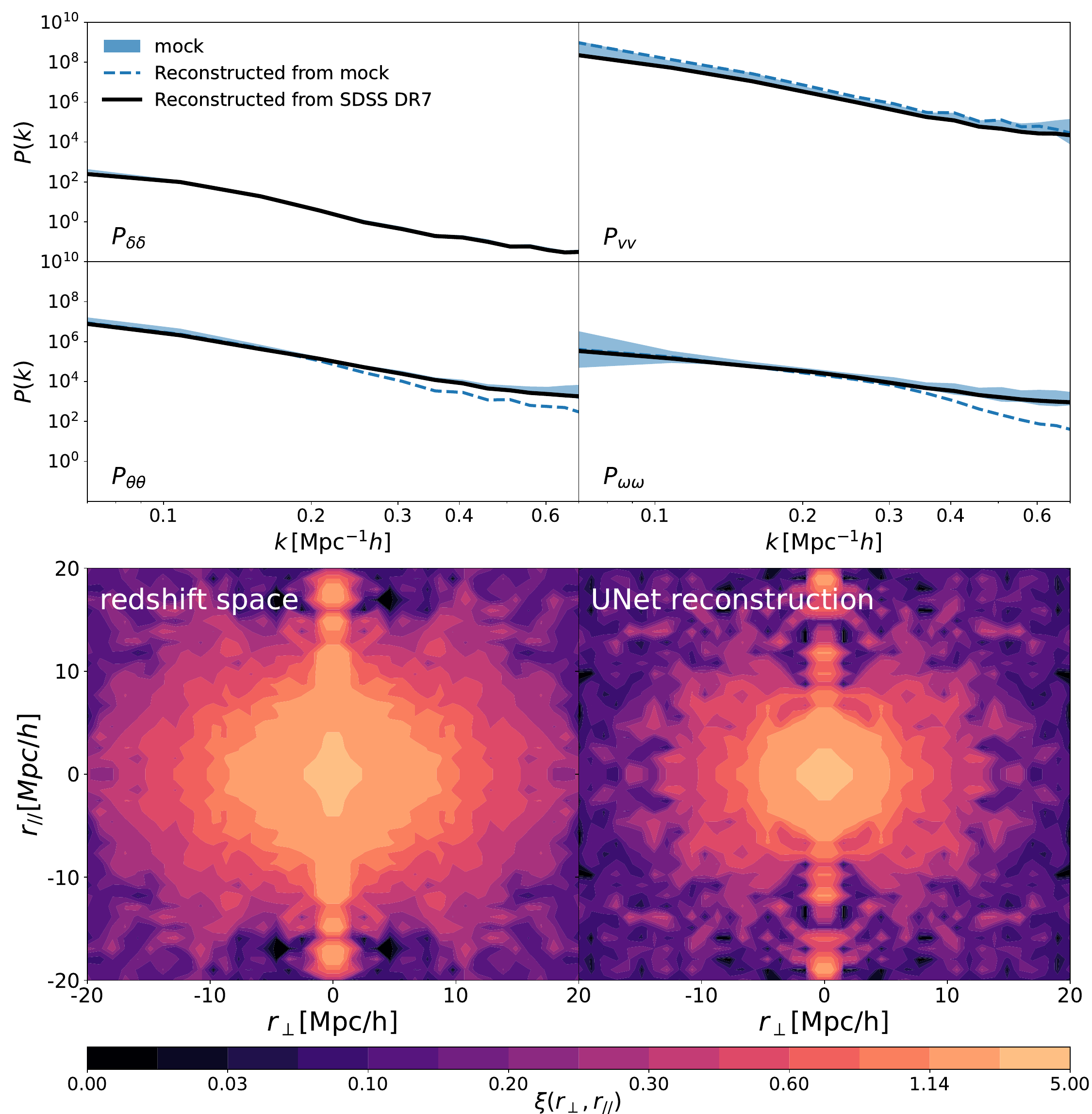}
\caption{Power spectrum and correlation function analysis. Top row shows four power spectra: density $P_{\delta\delta}$,  velocity $P_{vv}$,  divergence $P_{\theta\theta}$, and vorticity $P_{\omega\omega}$. Each displays mock data (blue shaded, $1\sigma$ uncertainty), mock reconstructions (blue dashed), and SDSS DR7 reconstructions (black solid). Bottom row shows 2D correlation functions $\xi(r_{\perp}, r_{\parallel})$: redshift-space (left) and UNet reconstruction (right). All the model and results are based on the density field evaluated with DTFE method.}
\label{fig:fig3}
\end{figure*}

We evaluate consistency with the concordance $\Lambda$CDM model using complementary frequency- and configuration-space diagnostics (Fig.~\ref{fig:fig3}). The upper panels show DTFE-based power spectra for the density ($P_{\delta\delta}$), velocity ($P_{vv}$), divergence ($P_{\theta\theta}$), and vorticity ($P_{\omega\omega}$) fields. Dotted curves denote the true simulated fields, shaded bands indicate the $\pm1\sigma$ spread across ten mock reconstructions, and solid black lines present the observational results. Density and velocity fields are smoothed with Gaussian kernels of $(4\,h^{-1}\,\mathrm{Mpc})^3$ and $(8\,h^{-1}\,\mathrm{Mpc})^3$, respectively (see Methods, section \textit{Data Pre-processing}). Fig.~\ref{fig:app_fig6} reports the analogous spectra for the CIC-based reconstructions, where both fields are smoothed with $(8\,h^{-1}\,\mathrm{Mpc})^3$ kernels.

The CIC-based measurements agree with the mock truths across the full range of scales, and the SDSS reconstructions remain comfortably within the quoted uncertainties. DTFE results display modest small-scale deviations ($k \gtrsim 2.5\,h\,\mathrm{Mpc}^{-1}$) in $P_{\theta\theta}$ and $P_{\omega\omega}$, plausibly reflecting the method’s heightened sensitivity to small-scale noise and structural complexity. Taken together, these comparisons attest to the fidelity of the AI-reconstructed fields and underscore their consistency with $\Lambda$CDM expectations.

\begin{figure*}
\centering
\includegraphics[width=1\textwidth]{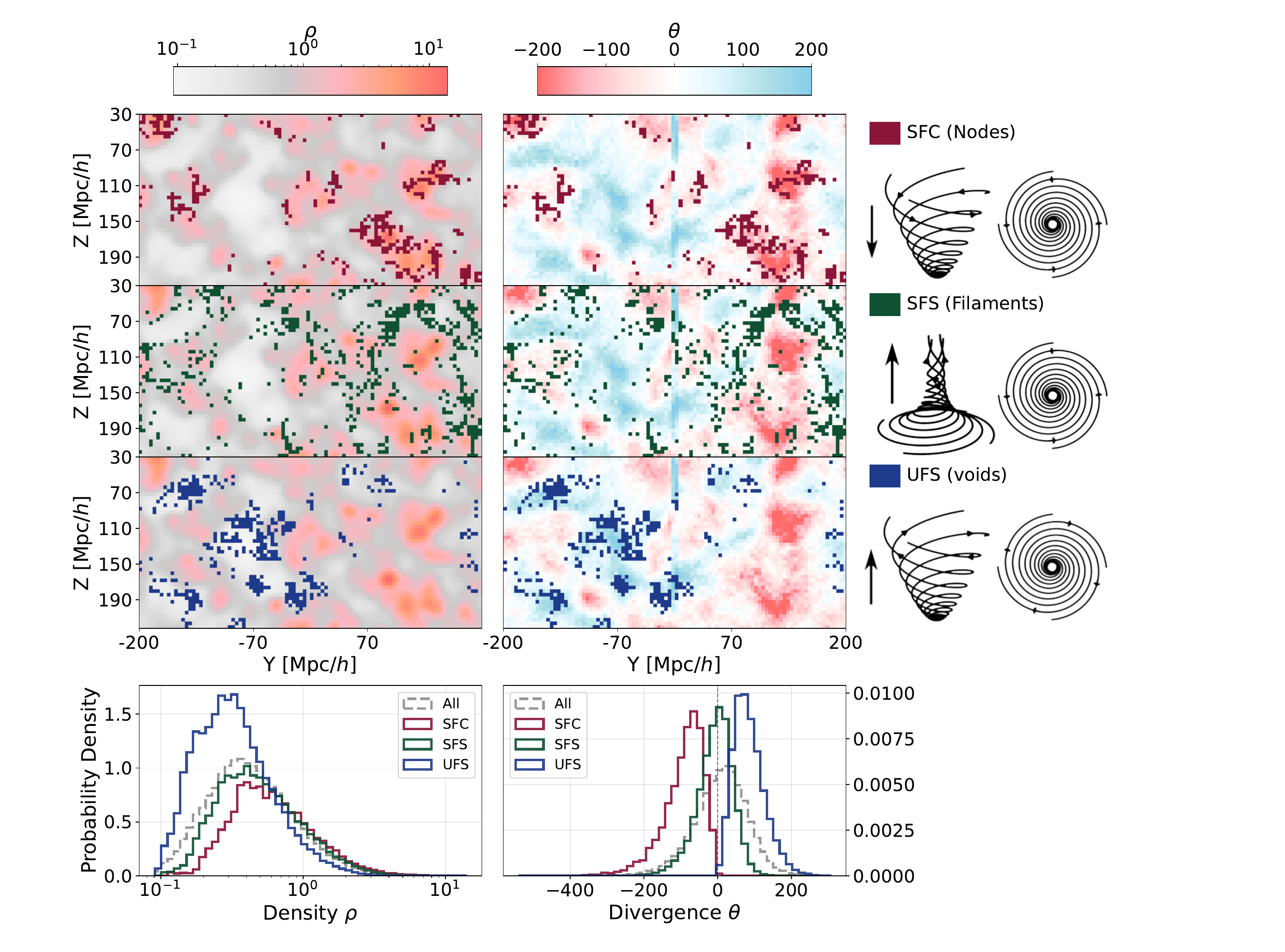}
\caption{The main $Y$--$Z$ slice shown for three rotational flow morphologies classified following \cite{WX2014ApJ} and \cite{Luo2023MNRAS}: SFC (stable focal compression, node-like regions), SFS (stable focal stretching, filamentary flows), and UFS (unstable focal stretching, void-like regions), from top to bottom. The backgrounds of left column plots are density field $\rho$, and reconstructed velocity divergence field $\theta$ in middle column plots. In these two columns, only grid cells belonging to the displayed morphology are overplotted as dark patches on top of the continuous fields. Right column illustrates the local velocity-flow patterns corresponding to SFC, SFS, and UFS. The bottom panels show probability density functions of $\rho$ (left) and $\theta$ (right) for all grid cells (black dashed) and for SFC, SFS, and UFS cells.}
\label{fig:fig4}
\end{figure*}

We further assess the reconstructed vorticity field through the asymmetric velocity-gradient tensor \cite{WX2014ApJ, Luo2023MNRAS}, which classifies local flow morphologies associated with the cosmic web. The spatial distribution of the four categories—SFC (dense nodes), SFS (filaments), UFC (walls), and UFS (voids)—agrees with expectations from the observed density field (Fig.~\ref{fig:fig4}). This concordance demonstrates that the model recovers not only statistical moments but also higher-order spatial coherence, reinforcing the conclusion that the AI reconstruction captures the essential dynamics encoded in $\Lambda$CDM-based simulations.

\subsection{Implications and outlook}
The lower panels of Fig.~\ref{fig:fig3} display the anisotropic two-point correlation function, $\xi(r_{\perp}, r_{\parallel})$, measured from the original SDSS catalogue (left) and from the reconstructed field after redshift-space distortion correction (right; see Methods, section \textit{Point-based catalog reconstruction}). The observed catalogue exhibits the expected anisotropies: large-scale Kaiser compression and small-scale Finger-of-God elongation. Both signatures are largely removed in the reconstructed catalogue, yielding a nearly isotropic signal characteristic of real-space clustering. The ability of the AI-derived velocity field to restore galaxy positions is essential for baryon acoustic oscillation analyses and Alcock–Paczynski tests. Because the network was trained solely on $\Lambda$CDM simulations, the suppression of both large- and small-scale redshift-space distortion features in actual data further indicates that the underlying dynamics are consistent with $\Lambda$CDM.

\section{Conclusion} \label{sec:conclusion}

In summary, we have reconstructed the vorticity field of cosmic peculiar velocities using an AI framework trained on realistic SDSS mock catalogues. The convergence of power-spectrum, correlation-function, and morphological diagnostics demonstrates that the reconstructed fields agree with $\Lambda$CDM predictions, eliminate redshift-space distortions, and reproduce the cosmic web traced by the density field. These results provide an independent validation of the concordance cosmological model and showcase the transformative potential of AI in observational cosmology.

Although preliminary, our reconstruction pipeline opens a path to studying nonlinear structure formation through high-fidelity velocity and vorticity measurements. The methodology can aid efforts to model or mitigate redshift-space distortions, characterise the kinematic Sunyaev–Zel'dovich effect, refine galaxy-clustering analyses, map the cosmic web, and probe galaxy–environment interactions. With forthcoming stage-IV surveys such as DESI, Euclid, Roman, PFS, and CSST, AI-based reconstructions can be integrated into data-processing pipelines to deliver unprecedented insight into cosmic evolution and fundamental physics. Together, these advances demonstrate that AI-enabled reconstructions can furnish the first observational view of cosmic vorticity while checking the validity of the concordance paradigm.

\appendix

\section{CIC Fields of SDSS Great Wall}
The main text focuses on DTFE-based reconstructions to illustrate the dynamical structures associated with the SDSS Great Wall. Fig.~\ref{fig:app_fig6} presents the corresponding results using CIC interpolation. The colour map shows galaxy number density in a $204 \times 408 \times 24~h^{-3}\mathrm{Mpc}^3$ slice, while white arrows denote velocity or vorticity vectors. The CIC reconstruction recovers coherent flows and vortical features similar to those obtained with DTFE, confirming that our conclusions do not depend sensitively on the chosen interpolation scheme.

\section{Point-based catalog reconstruction}
To mitigate the loss of small-scale information incurred by gridding, we developed a procedure to reconstruct field values at galaxy positions. Algorithm~\ref{alg:algo1} outlines the method. For a field defined on $m$ grid cells of size $d\,h^{-1}\,\mathrm{Mpc}$, each cell is subdivided into $b$ subcells. By shifting the grid by $d/n\,h^{-1}\,\mathrm{Mpc}$ along each Cartesian axis, we generate $n^3$ offset versions of the field, yielding distinct samples for each original cell. The field value at an arbitrary point is approximated as a weighted combination of these $n^3$ samples.

Empirically we find that $n=10$ suffices to reconstruct point-cloud velocities with the desired accuracy. During training, we randomly shift the grid origin along each axis up to $t\leq 10$ times, each shift being $d/n\,h^{-1}\,\mathrm{Mpc}$. This augmentation encourages the UNet to learn multiple velocity representations per grid cell and improves fidelity when mapping back to discrete galaxy positions.

\refstepcounter{algorithm}
\begin{figure}
\centering
\fbox{
\parbox{0.92\linewidth}{
\small
\textbf{Algorithm 1: Point cloud reconstruction (1D)}\\

\small
\begin{tabbing}
\hspace{1em}\=\hspace{1em}\=\kill
\textbf{Input:} $CIC(v, boxsize, ngrid, x_{\min}, x_{\max})$ \\
\textbf{Require:} $x_{\min} < x_0 < x_{\max}$ \\
$v_0 \leftarrow 0$ \\
$N \leftarrow n$ \\
\textbf{while} $N \neq 0$ \textbf{do} \\
\> $x_{\min}, x_{\max} \leftarrow x_{\min}+d/n,\; x_{\max}+d/n$ \\
\> $i \leftarrow (x_0 - x_{\min}) / d$ \\
\> $I_N \leftarrow (x_0 - x_{\min} - i d)\, n / d$ \\
\> $v_N \leftarrow CIC(v, d m, m, x_{\min}, x_{\max})[i]$ \\
\> $N \leftarrow N - 1$ \\
\textbf{end while} \\
$N \leftarrow n$ \\
\textbf{while} $N \neq 0$ \textbf{do} \\
\> $i \leftarrow I_N$ \\
\> $v_0 \leftarrow v_0 + \frac{1}{i} v_i^{\, i}$ \\
\> $N \leftarrow N - 1$ \\
\textbf{end while}
\end{tabbing}
}
}
\label{alg:algo1}
\end{figure}

\section{Kinematic Morphology}
Following \cite{2014ApJ...793...58W}, we characterise local flow morphology using the velocity-gradient tensor
\begin{eqnarray}
A_{ij}=\frac{\partial u_i}{\partial x_j},
\end{eqnarray}
where $\mathbf{u}$ is the velocity field. In the V-web formalism the symmetrised tensor
\begin{equation}
\mathbf{A}^{\mathrm{s}}=\frac{1}{2}\left(\mathbf{A}+\mathbf{A}^{\mathsf{T}}\right)=\mathbf{R}^{-1} \operatorname{diag}\left(\lambda_1, \lambda_2, \lambda_3\right) \mathbf{R}
\end{equation}
has three real eigenvalues $\lambda_i$, with $\mathbf{R}$ denoting the rotation matrix. Comparison with a threshold $\lambda_{\mathrm{th}}$ allows classification into nodes (all $\lambda_i < \lambda_{\mathrm{th}}$), filaments (two eigenvalues below the threshold), sheets (one below), or voids (all above).

Alternatively one can use the rotational invariants $s_i=\{s_1,s_2,s_3\}$ of $A_{ij}$, defined through the characteristic equation $\det(\mathbf{A}-\lambda\mathbf{I})=0$:
\begin{equation}
s_1=-\operatorname{tr}[\mathbf{A}], \qquad s_2=\frac{1}{2}\left(s_1^2-\operatorname{tr}\left[\mathbf{A}^2\right]\right), \qquad s_3=-\det[\mathbf{A}].
\label{eq:coefficients}
\end{equation}
These coefficients remain real even when shell-crossing generates asymmetric tensors with complex eigenvalues, providing a continuous description of the flow.

In canonical form an asymmetric tensor can be written as
\begin{equation}
\mathbf{A}=\mathbf{R}^{-1}\left(\begin{array}{ccc}
a & -b & 0\\
b & a & 0\\
0 & 0 & c
\end{array}\right) \mathbf{R},
\label{eq:asymmetric}
\end{equation}
with $a$, $b$, and $c$ real and eigenvalues $\lambda_{1,2} = a \pm ib$ and $\lambda_{3} = c$. The local trajectory obeys
\begin{equation}
\frac{d \mathbf{x}}{d \tau}=\mathbf{A} \mathbf{x}.
\end{equation}
The complex eigenvalues span a plane in which trajectories follow logarithmic spirals
\begin{equation}
r=r_0 e^{m \alpha},
\end{equation}
where $(r,\alpha)$ are polar coordinates, $m=a/b$ controls the pitch angle, and $r_{0}$ depends on initial conditions. Large $|m|$ produces gently winding spirals, whereas small $|m|$ yields tightly wound trajectories. Motion perpendicular to the plane is governed by $\lambda_3=c$.

\begin{extdatafigure}
\centering
\includegraphics[width=\textwidth]{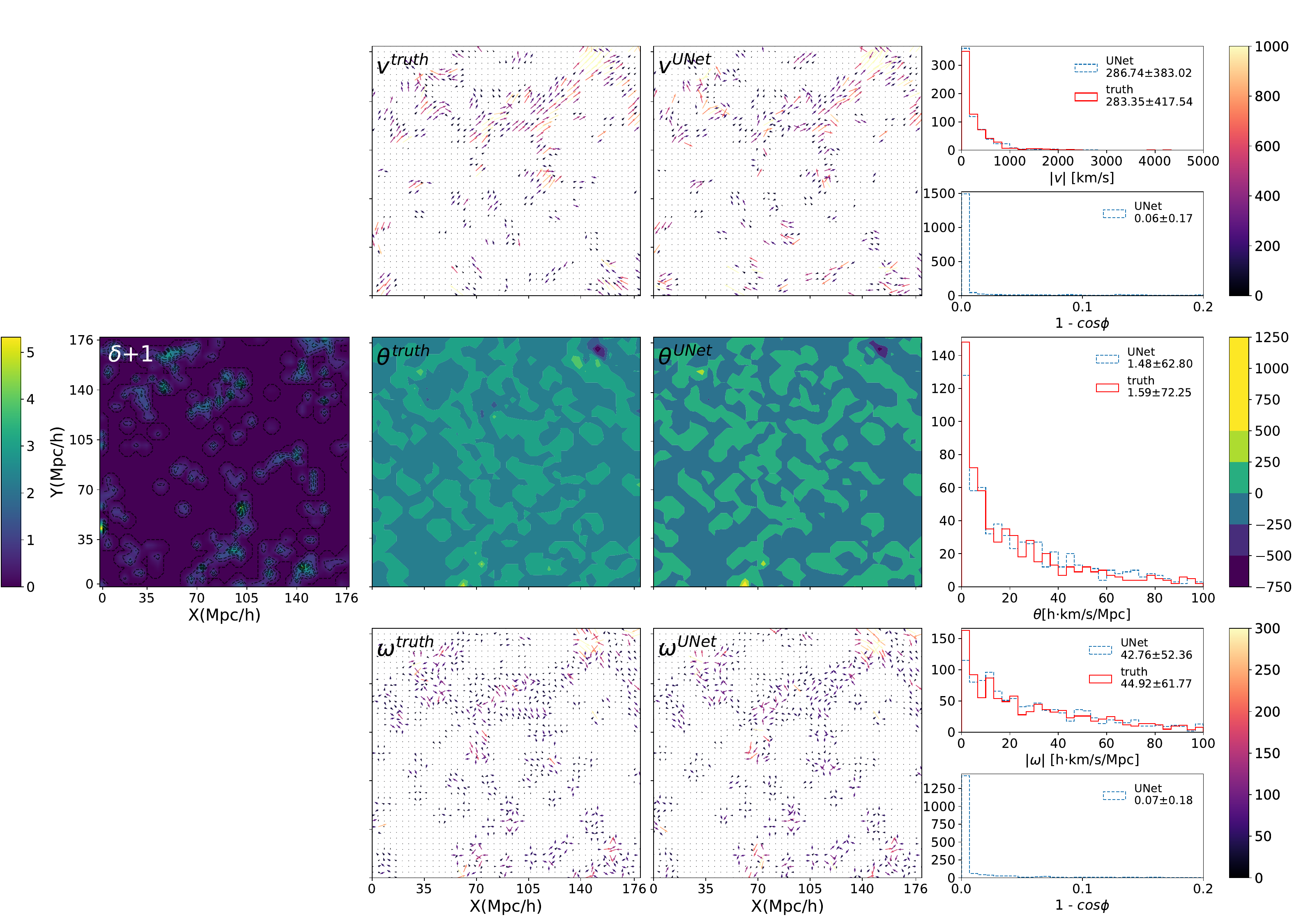}
\caption{Pixel-wise comparison between the UNet-reconstructed and true velocity fields using the SDSS mock test set. From top to bottom, we show slices of the velocity, divergence, and vorticity fields (volume: $122^3\,(h^{-1}\mathrm{Mpc})^3$). From left to right, the panels display the galaxy number density field, the UNet-reconstructed field, the true field, and their corresponding histograms. Colored arrows indicate vector magnitude and direction. The histograms show the distributions of field magnitudes as well as the angular differences between the true and reconstructed velocity and vorticity vectors. The UNet robustly recovers the velocity, divergence, and vorticity structures, with strong visual agreement and statistical consistency in the reconstructed fields.}
\label{fig:app_fig4}
\end{extdatafigure}

Excluding degenerate cases, the signs of $m$ and $c$ determine the four canonical focal-flow classes of \cite{2014ApJ...793...58W}. Inward spirals ($m<0$) with $c<0$ define the stable focal compression (SFC) morphology, characteristic of node-like regions. When $m<0$ but $c>0$, the flow still spirals inward while stretching along the transverse direction, producing the stable focal stretching (SFS) morphology that traces filaments. Outward spirals ($m>0$) with $c<0$ correspond to unstable focal compression (UFC), associated with sheet- or wall-like environments, whereas $m>0$ and $c>0$ yield unstable focal stretching (UFS), which marks void-like, expanding regions. These assignments are consistent with both the invariant thresholds and the spatial clustering observed in our reconstructions.

\section{Pixel-wise Check}
To assess accuracy and robustness we compare the UNet-predicted halo velocity field with the simulation truth on a voxel-by-voxel basis. Fig.\ref{fig:app_fig4} displays a representative slice from the mock test set, illustrating the galaxy number density together with the reconstructed velocity, divergence, and vorticity fields within a $122 \times 122 \times 122~(h^{-1}~\mathrm{Mpc})^3$ volume for the CIC-based reconstructions. For each quantity the true and reconstructed fields are plotted side by side. Coloured arrows indicate the projected mean velocity in each cell, with direction and length encoding orientation and speed, and colour depicting magnitude.

As expected, regions of high galaxy density host stronger flows driven by gravitational collapse. The reconstructed velocity field is visually indistinguishable from the truth, illustrating the network’s ability to capture both large-scale morphology and small-scale features even in sparsely sampled regions. Histograms of velocity magnitudes and directions (right-hand panels) quantify the agreement: the reconstructed magnitudes, $286.74 \pm 283.02~\mathrm{km}~\mathrm{s}^{-1}$, match the truth to within 98\%, and the mean directional discrepancy is $1 - \cos \phi = 0.06 \pm 0.17$.

The bottom panels show the vorticity field $\boldsymbol{\omega}$, which peaks in nonlinear, high-density environments and declines rapidly elsewhere. Despite the difficulty of inferring vorticity from sparse tracers, the reconstruction achieves a residual $|\omega^{\rm true} - \omega^{\rm U\text{-}Net}| = 2.16 \pm 9.41~h~\mathrm{km}~\mathrm{s}^{-1}~\mathrm{Mpc}^{-1}$, and the angular error distribution has a mean $\Delta \phi \approx 21.6^{\circ} \pm 34.9^{\circ}$. The divergence field exhibits similarly strong agreement, with $|\theta^{\rm true} - \theta^{\rm U\text{-}Net}| = 0.11 \pm 9.45~h~\mathrm{km}~\mathrm{s}^{-1}~\mathrm{Mpc}^{-1}$, confirming that the model accurately reproduces multiple dynamical diagnostics.

\begin{extdatafigure}
\centering
\includegraphics[width=\textwidth]{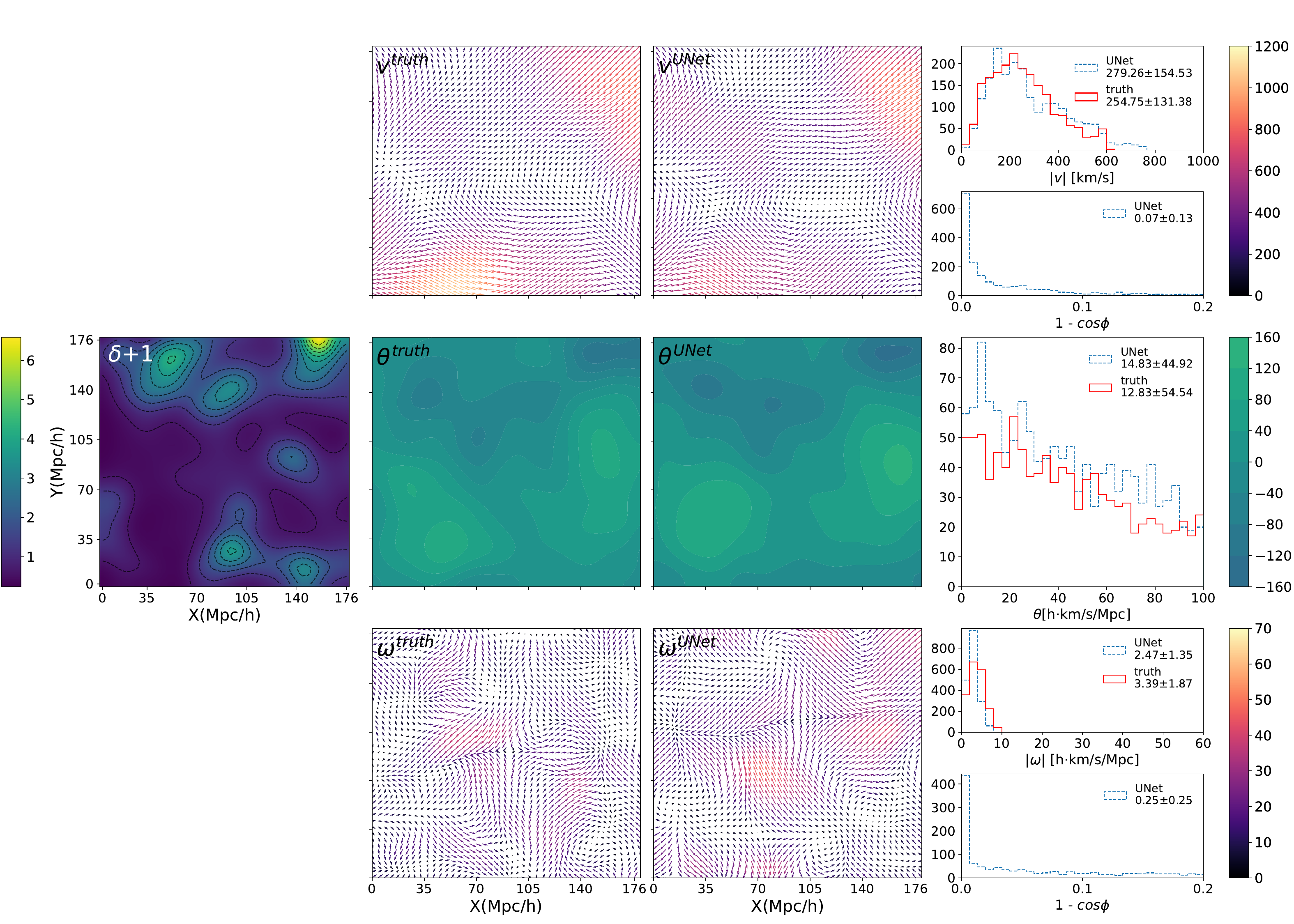}
\caption{Same as Fig.~\ref{fig:app_fig4}. Pixel-wise comparisons between the UNet-reconstructed and true velocity fields for the DTFE-based reconstruction using the SDSS mock test set. An 8 $h^{-1}\rm{Mpc}$ Gaussian smoothing kernel is applied to the DTFE field to mitigate spurious triangular interpolation noise.}
\label{fig:app_fig5}
\end{extdatafigure}

Fig.~\ref{fig:app_fig5} shows the corresponding plot for the DTFE-based reconstruction. On large scales, the reconstructed fields are nearly identical to the ground truth, while on small scales they exhibit mild but noticeable deviations. This behavior is consistent with the power-spectrum comparison and likely reflects the more complex morphology produced by DTFE together with the limited size of the training set. Achieving more accurate small-scale structures will require additional training data. 

\acknowledgments

This work was supported by the National Natural Science Foundation of China (12373005, 12473097, 12073088), the National SKA Program of China (2020SKA0110401, 2020SKA0110402, 2020SKA0110100), the National Key R\&D Program of China (2020YFC2201600), the China Manned Space Project, the Fundamental Research Funds for the Central Universities, Sun Yat-sen University(No. 24qnpy122), the Basic and Frontier Research Project of PCL (2025QYB012) and the Guang-dong Basic and Applied Basic Research Foundation (2024A1515012309). We also acknowledge the Beijing Super Cloud Center (BSCC) and Beijing Beilong Super Cloud Computing Co., Ltd. (http://www.blsc.cn/) for providing HPC resources that substantially supported this study. We thank Qiufan Lin for helpful discussions. 


\bibliographystyle{JHEP}
\bibliography{biblio.bib}

\end{document}